\documentclass[showpacs,amsmath,amssymb,prd,twocolumn]{revtex4}
\usepackage{amssymb}

\usepackage{latexsym}
\usepackage{graphicx}
\usepackage{slashed}
\usepackage{pifont}
\usepackage{color}

\newcommand{\req}[1]{Eq.\,(\ref{#1})}
\newcommand{\beqn}{\begin{equation}}
\newcommand{\eeqn}{\end{equation}}
\newcommand{\gsgg}{g_{\sigma\gamma\gamma}}
\newcommand{\mpl}{m_{\mathrm{P}}}

\newcommand{\tr}{\mathrm{tr}\,}

\begin{document}
\title{Electromagnetic signal of the QCD phase transition in neutron star mergers}
\author{Pisin Chen$^{1,2,3}$ and Lance Labun$^{1}$}
\affiliation{$^1$ Leung Center for Cosmology and Particle Astrophysics (LeCosPA)\\
National Taiwan University, Taipei, 10617 Taiwan\\
$^2$ Department of Physics and Graduate Institute of Astrophysics\
National Taiwan University, Taipei, 10617 Taiwan\\
$^3$ Kavli Institute for Particle Astrophysics and Cosmology\\
SLAC National Accelerator Laboratory\\
Menlo Park, CA 94025, USA}

\date{May 30, 2013} 

\begin{abstract} 
Mergers of binary neutron stars create conditions of supranuclear density $n\gtrsim n_{\rm nuc}\simeq 0.17~{\rm fm}^{-3}$ and moderate temperature $50\lesssim T \lesssim 90~{\rm MeV}$.  These events thus probe a sensitive region of the density-temperature phase diagram of QCD matter.  We study photon production by the QCD conformal anomaly for a signal of a possible transition to quark degrees of freedom during the merger.  We discuss energy loss due to photon radiation as a cooling mechanism that is sensitive to the bulk viscosity and thermal conductivity of the quark matter.
\end{abstract}

%97.60.Jd Neutron stars (see also 26.60.-c Nuclear matter aspects of neutron stars in-Nuclear physics)
%26.50.+x Nuclear physics aspects of novae, supernovae, and other explosive environments 
%21.65.Qr Quark matter (see also 12.38.Mh Quark-gluon plasma in quantum chromodynamics; 25.75.Nq Quark deconfinement, quark-gluon plasma production and phase transitions in relativistic heavy-ion collisions)
%98.70.Rz 	gamma-ray sources; gamma-ray bursts 
% --
%26.60.-c Nuclear matter aspects of neutron stars

\maketitle
%%%%%%%%%%%%%%%%%%%%%%%%%%%%%%%%%%%%%%%%%%%%%%%%%%%%%%%%%%%%
\section{Introduction}

Neutron stars offer one of the few options to obtain observational data on the properties of bulk nuclear matter at and above the saturation density $n_{\rm nuc}\simeq 0.17\:{\rm fm}^{-3}$.  Dense nuclear matter and the expected transition to quark and gluon degrees of freedom at higher density are challenging to study also on the theoretical side. 
One of the challenges met is that a single neutron star's mass, radius and moment of inertia tend to be only weakly sensitive to the properties of the matter in the interior and many phenomenologically quite distinct equations of state support the observations~\cite{Lattimer:2006xb}.  Dynamical effects, such as the neutrino cooling rate, can offer additional information~\cite{Page:2005fq,Heinke:2010cr}.

Neutron star mergers will also yield information about dense nuclear matter.  Binary neutron stars (bNS) are expected to be a significant population in galaxies~\cite{Kim2003,LIGO2010} and merger events are thought to be the sources of short gamma ray bursts (sGRB).  The bNS-sGRB connection could be confirmed if a signal within the sGRB, specific to bNS mergers, is identified.  This is one of the reasons gravitational waves from a bNS merger are sought by LIGO and Virgo~\cite{LIGO2012}, and numerical simulations are used to study the waveform of a binary merger signal for its dependence on the characteristics of the stars involved~\cite{LRRFaber} and the properties of the dense matter inside the stars~\cite{Bauswein:2009im,Bauswein:2011tp,Bauswein:2012ya,Sekiguchi:2011mc,Andersson:2009yt}.

We propose to complement these efforts by studying electromagnetic signals from the bNS merger and seeking a signature sensitive to the properties of the dense matter.  To this end, we note that a collision of neutron stars is semirelativistic ($v\simeq 0.3c$) and the density is near to the nuclear saturation also in the outermost layers of each star.  These facts create conditions during the merger in which it is possible, even likely, that the degenerate gas of protons and neutrons undergoes a phase transition to a state dominated by quark (and perhaps gluon) degrees of freedom.  This possibility has been invoked before as the basis of a model to explain sGRBs~\cite{Chen:2013tp}. Different processes are important in the new phase, and among them we may find a phenomenon characteristic of the transition having occurred, just as enhancement of strange-flavored baryons is an important signal of the creation of quark-gluon plasma (QGP) in laboratory experiments~\cite{Koch:1986ud,Rafelski:2011ek}.  

In this first study, the process we focus on is photon production due to the conformal anomaly of quantum chromodynamics (QCD).  The conformal anomaly is the non-conservation of the dilatational current $S^{\mu}$ due to the violation of scale invariance at the quantum level even in the limit of massless quarks, see~\cite{Ellis:1970yd,Chanowitz:1972vd} for discussion.  The source of the dilatational current is the trace of the energy-momentum tensor, $\partial_\mu S^{\mu}=T_{\mu}^{\mu}$, whose correlators correspond to propagation of bulk modes in the plasma.  Bulk modes acquire an effective coupling to photons through a quark loop (the anomaly), because quarks also carry electric charge.  

An important reason to study the conformal anomaly is that it can be substantially enhanced near a phase change.  This is seen in QCD on the lattice, evaluating $T_{\mu}^{\mu}/T^4=(\varepsilon-3P)/T^4$ for low density, finite temperature QGP, see for example Fig. 1 of~\cite{Bazavov:2009zn}.  Lattice studies extended to finite density suggest the peak in $T^{\mu}_{\mu}$ increases with chemical potential~\cite{Basak:2009uv}, and model studies of high density matter show a related phenomenon in a peak in the bulk viscosity near to the transition~\cite{Li:2009by}.  Thus, although the conformal anomaly is present in both hadron and quark phases, its enhancement and consequent photon production near the phase change may serve as a signal of transition.

It was recently pointed out that the QCD conformal anomaly could be a source of photons in the QGP produced in collider experiments~\cite{Basar:2012bp}.  We extend the analysis of this effect to the possibility of a quark phase arising during bNS mergers.  An essential difference from the QGP formed in the lab is that high density plays an important role in the possible transition, and the properties of the hot, high density quark matter are less well known.  In particular, we do not know the location of a possible hadron-quark transition in the density-temperature plane.  We shall show that, during the bNS merger, even matter near the surfaces of the stars enters the sensitive region of the phase diagram where a transition may be expected.  

The time scale of QCD processes is much shorter than the weak interaction and mechanical time scales of the merger:
\beqn\label{times}
\!\!\!\tau_{\rm QCD}\lesssim 10^{-20}{\rm \,s}\ll \tau_{\rm weak}\sim 10^{-7}\!-10^{-6}{\rm \,s}\ll \tau_{\rm merger}\sim 10^{-3}{\rm \,s}.
\eeqn 
The relevant $\tau_{\rm QCD}$ is the relaxation time scale arising from bulk transport properties and is one of the outcomes of the present study. We will show that the time scale due to cooling by photon emission is much shorter than the weak interaction time scale, see Eqs.~\eqref{tauE} and \eqref{tauk} below. The weak interaction relaxation time scale $\tau_{\rm weak}$ arises from consideration of Urca processes and will clearly be relevant over the long duration of the merger, see e.g.~\cite{Page:2005fq}.  $\tau_{\rm merger}$ is the orbital timescale of coalescing binary, and numerical simulations indicate that the system relaxes toward its final state over 5-10$\tau_{\rm merger}$.

To set the stage and motivate the claim that a QCD phase transition is probable during a merger of neutron stars, we first estimate the temperature and density in the domains where the stars are coming into contact.  We work in units with $\hbar=c=1$ throughout.

%\vskip0.2cm
%%%%%%%%%%%%%%%%%%%%%%%%%%%%%%%%%%%%%%%%%%%%%%%%%%%%%%%%%%%%%%%%%%%%%%%%%
\section{Analytic Model for Conditions of Merger}\label{sec:kinematics}

Determining the kinematics and local conditions in the stars up to and during the merger is in general a hard problem, due to the interplay of strong field general relativity and dense nuclear matter.  Much effort is devoted to numerical simulations incorporating nuclear equations of state in the general relativistic dynamics of the binary inspiral and merger\cite{Bauswein:2009im,Bauswein:2011tp,Bauswein:2012ya,Sekiguchi:2011mc}.  Here, we follow in the spirit of Shapiro~\cite{Shapiro:1998sy} and make analytic estimates of the kinematics at the onset of the collision in order to determine local temperatures and densities that can be achieved.  The temperature will be estimated from the amount of kinetic energy that must be dissipated for the stars to merge into a single object, and the density by conservation of baryon number in the volume where the matter from the two stars combine.

In the center-of-mass frame of the binary system, the relative distance of the stars is denoted $r$ and the corresponding radial velocity, or rate of closure, $v_r$. We consider the collision to begin when the surfaces of the two neutron stars are expected to ``touch,'' that is, when the relative distance is twice the radius of the stars
\beqn
r_{\rm coll}\simeq 2R_*\simeq 20~{\rm km}
\eeqn
putting in a typical expected value for $R_*$~\cite{Lattimer:2006xb}.  

The kinetic energy dissipated is due to the radial velocity, $v_r$, which we evaluate at the onset of collision.  We start from the total energy of the binary system, defined as the sum of kinetic $K$ and potential $V$ energies.  The total energy with post-Newtonian corrections [Eq.\,(194) of~\cite{LRRBlanchet}] is
\begin{align}\label{infallE}
K\!+\!V=E=&-\frac{M\nu x}{2}\left[1-\frac{3}{4}\left(1+\frac{\nu}{9}\right)x \right.\\ \notag
&\hspace*{14mm}\left.-\left(\frac{27}{8}-\frac{19}{8}\nu+\frac{\nu^2}{12}\right)x^2+{\mathcal O}(x^3)\right] 
\end{align}
Here $M=m_1+m_2$ is the total mass of the binary system, and $\nu=m_1m_2/M^2$ is the symmetric mass ratio, which varies between $0,1/4$ achieving the maximum for $m_1=m_2$. The frame-invariant post-Newtonian expansion parameter,
\beqn
x=(GM\omega)^{2/3}\sim {\mathcal O}(v^2)
\eeqn
is related to $GM/r$ by [Eq.\,(193) of~\cite{LRRBlanchet}]
\beqn\label{PNexps}
\frac{GM}{r}=x\left[1+\left(1-\frac{\nu}{3}\right)x+\left(1-\frac{65}{12}\nu\right)x^2+\mathcal{O}(x^3)\right]
\eeqn
up to second post-Newtonian order.  For $r_{\rm coll}=20~{\rm km}$, $M=3M_{\odot}$ (solar mass) and $\nu=1/4$, we have $x_{\rm coll}=0.21$.

The total energy \req{infallE} can be given coordinate-invariant meaning~\cite{LRRBlanchet}, but a splitting between kinetic and potential contributions has not been uniquely given in the post-Newtonian framework.
Using the moment of inertia of two point masses, we define the kinetic energy to have the form
\begin{align}\label{Ttot}
K&=\frac{1}{2}M\nu v_r^2 + \frac{1}{2}M\nu r^2\omega^2 \\ \notag
 &=\frac{1}{2}M\nu v_r^2+\frac{1}{2}M\nu x\left[1-2x\left(1-\frac{\nu}{3}\right) \right. \\
\notag &\hspace*{25mm} \left.+\left(1+\frac{53}{6}\nu+\frac{\nu^2}{3}\right)x^2+...\right]
\end{align}
In the second line, we have used \req{PNexps} to write $r^2\omega^2$ only in terms of $x$ and carried out the post-Newtonian reexpansion in $x$.

To determine the kinetic energy and solve for the squared radial velocity, we need an expression for the Newtonian potential. The two natural choices $V\simeq -M\nu x$ and $V=-GM^2\nu/r$ are equivalent at Newtonian level but differ significantly with the inclusion of post-Newtonian corrections. Putting each form in for $V$ and setting the remainder equal to the kinetic energy \req{Ttot}, we obtain
\beqn\label{vrest}
v_r^2\simeq
\begin{cases}
 \left(\frac{11}{4}-\frac{7\nu}{12}\right)x^2 \quad &  \mathrm{for}~~V\simeq -M\nu x \\
 \left(\frac{19}{4}-\frac{5\nu}{12}\right)x^2 \quad &  \mathrm{for}~~V\simeq -GM^2\nu/r
\end{cases}
\eeqn
For $x=x_{\rm coll}$ and $\nu=1/4$, we find $v_r^2=0.11$ for the first definition of the potential and respectively $v_r^2=0.20$ for the second.

In the above formulae, the masses appearing are the gravitational masses.  Taking into account the gravitational defect which is 8--12\% depending on the neutron star equation of state, the total system mass corresponds to a baryonic mass of $M_b=M\left(1+\frac{\delta M}{M}\right)\sim (1.1\pm 0.02)M$.
For this range of mass defects, the energy per nucleon released in the collision is 
\begin{align}\label{EperA}
T&\simeq\frac{1}{2}\frac{m_N}{1+\delta M/M}v_r^2 \\ \notag
 &\simeq 
\begin{cases}
48\pm 1~{\rm MeV} \quad &  \mathrm{for}~~V\simeq -M\nu x \\
85\pm 2~{\rm MeV} \quad &  \mathrm{for}~~V\simeq -GM^2\nu/r
\end{cases}
\end{align}
using respective $v_r$ from \req{vrest}, the nucleon mass $m_N\simeq 939\:{\rm MeV}$ and dropping the small dependence on the mass ratio $\nu$.

Designed to be conservative, the estimate $T\simeq 50~{\rm MeV}$ is at the low end of the range of the thermal energy seen in some numerical simulations, e.g.~\cite{Bauswein:2009im}. Our outcome is compatible with Shapiro's estimates~\cite{Shapiro:1998sy} that the remnant object achieves quasiequilibrium at a temperature of order $140~{\rm MeV}$.  In agreement with Shapiro, we also check that the radiated gravitational wave energy, which decreases the kinetic energy at collision, is a negative correction similar to or smaller than positive corrections such as dissipation of angular momentum during the collision. Taken together, these analytic and numerical results imply that temperatures in the range $T=50-90~{\rm MeV}$ are achieved in the course of the collision.

To estimate density, we start from the fact that the baryon density near the surface of an isolated neutron star in equilibrium is approximately $n_{\rm crust}/n_{\rm nuc}=0.15\!-\!0.6$.  (One can check these by a simple argument for the binding energy of a nucleon at the star's surface.) Considering a superposition of baryon number in the volume where the stars overlap, we expect that the density achieved is 2 to 4 times the initial density, 
\beqn\label{nest}
 n/n_{\rm nuc}\sim (2\!-4\!)n_{\rm crust}/n_{\rm nuc}\sim 0.3\!-\!2.4.
\eeqn  
At zero temperature, for a degenerate gas of baryons, these densities correspond to baryon chemical potentials $\mu_B=\sqrt{(3\pi^2n)^{2/3}+m_N^2}=944\!-\!1240~{\rm MeV}$.

%\vskip0.2cm
%%%%%%%%%%%%%%%%%%%%%%%%%%%%%%%%%%%%%%%%%%%%%%%%%%%%%%%%%%%%%%%%%%%%%%%%%
\section{Nuclear matter at the onset of collision}\label{sec:nucmatter}

The temperature and density ranges [\req{EperA} and \req{nest}] point to a region of the nuclear/QCD phase diagram that is challenging to study.  The upper limits of these ranges lie in a region of the phase diagram where quarks are expected degrees of freedom, according to recent reviews~\cite{Weise:2012yv,Fukushima:2013rx}.  Note that the lower end of the temperature estimate is higher than the end point of the expected first-order liquid-gas phase transition in nuclear matter~\cite{Kapusta:2006pm}, and we also do not expect to reach into the higher density--lower temperature domain where color superconducting phases become possible~\cite{Alford:2007xm}.  

Above we estimated the density in the surface layers because we are interested in prompt electromagnetic signals and we will see below that photons have a short mean-free path in the high density matter.  Only domains of the new phase that are near the surface produce electromagnetic radiation that can escape the star within the millisecond time scale of the merger.  
Because of the relatively low density $n<n_{\rm nuc}$ near the surfaces of the initial stars, this matter is most likely nuclear matter, including possible nuclear ``pasta'' configurations~\cite{Watanabe:2000rj,Sonoda:2007sx}.  Before the collision, the near-surface matter is expected to be ``normal'' neutron-rich matter without pion or kaon condensates or quark matter of any type.  

Weak interactions would bring strange quarks and strange-flavored hadrons into chemical equilibrium in either quark or hadronic phases.  By considering time scales shorter than the weak relaxation time $\tau_{\rm weak}$, we thus simplify the analysis to the presence of only light-flavored quarks and corresponding hadrons.  This simplification would not apply if the temperature is of the order the mass of strange quark $m_s\sim 95\pm 5~{\rm MeV}$ and gluonic degrees of freedom are also liberated in the transition, because in this case strangeness could achieve chemical equilibrium on QCD time scales by the same mechanism as in low-density QGP~\cite{Koch:1986ud,Rafelski:2011ek}.

Below, we will describe the environment assuming a first order transition occurs between the hadron- and quark-dominated phases, and consequently separation of the phases into domains.  Phase coexistence is supported by lattice studies that use the canonical ensemble~\cite{Alexandru:2005ix,deForcrand:2006ec,Li:2011ee}, though not all model studies and symmetry arguments find a first order transition.  While the photon production mechanism does not require the first order transition, we expect a first order transition could lead to large photon production because the jump in entropy would lead to high bulk viscosity, which is an important property entering the production rate, as seen below.  Also in the case the transition is second order or weaker, photon production can be large due to a peak in the bulk viscosity near the transition~\cite{Li:2009by}.  For reviews of the outcomes of various approximations and models of dense matter, see~\cite{Weise:2012yv,Fukushima:2013rx}. 

%\vskip0.2cm
%%%%%%%%%%%%%%%%%%%%%%%%%%%%%%%%%%%%%%%%%%%%%%%%%%%%%%%%%%%%%%%%%%%%%%%%%
\section{Photon Production by the Conformal Anomaly}

The conformal anomaly of QCD arises from the breaking of scale invariance by the quantum effects of running of the coupling and dimensional transmutation in addition to explicit breaking by quark masses.  The anomaly is expressed as the nonvanishing divergence of the dilatation current $S^{\mu}$, 
\beqn\label{confanom}
\partial_{\mu}S^{\mu}=T_{\mu}^{\mu}=\frac{\beta(\alpha_s)}{4\alpha_s}\tr[G_{\mu\nu}G^{\mu\nu}]+\sum_fm_f(1+\gamma_f)\bar q_fq_f,
\eeqn
where $T_{\mu}^{\mu}$ is trace of the energy-momentum tensor, $\beta(\alpha_s)$ is the renormalization group beta function of the QCD coupling $\alpha_s$ and $G^{\mu\nu}$ the gluon field strength tensor.  For each quark flavor $f$, $m_f$ is the mass and $\gamma_f$ is the anomalous dimension. As noted in the Introduction, the conformal anomaly [\req{confanom}] is nonzero in both hadron and quark phases, but we can expect it is enhanced where temperature and density are near or in the transition region between the phases.

The expectation value of the trace of the energy momentum tensor $\langle T^{\mu}_{\mu}\rangle$ depends on the density and temperature of the nuclear matter, and the dependence has been studied by several authors~\cite{Cohen:1991nk,Saito:2005rv}.  In the finite temperature and density expansion, one can always isolate the vacuum contribution [for example as in Eq. (3.6) of~\cite{Cohen:1991nk}], which arises from the quantum anomaly  breaking conformal invariance.  This anomalous part is due to ultraviolet physics, being given by loop corrections by virtual quark and gluons with corresponding momenta at QCD scale or higher.  In fact, in the low-energy effective theory we will use to describe the effects due to $T^{\mu}_{\mu}$,  contributions from all momenta are integrated, as noted also by~\cite{Basar:2012bp} including all six quark flavors when evaluating the anomaly.  In short, although the physical observable $\langle T^{\mu}_{\mu}\rangle$ depends on the density and temperature, the effective interaction leading to photon production is a quantum effect and determined by the vacuum part of the anomaly.

In the low-energy effective theory~\cite{Ellis:1984jv}, the dilatational current $S^{\mu}$ is described by a colorless scalar degree of freedom $\sigma$ such that~\cite{Basar:2012bp}
\beqn\label{sigmamatrixelem}
\langle 0|\partial_{\mu}S^{\mu}|\sigma\rangle = m_\sigma^2f_\sigma
\eeqn
with $m_\sigma$ the mass of the scalar mode and $f_\sigma$ the decay constant, analogous to $f_\pi$.  

\begin{figure}
\centerline{
\begin{picture}(100,60)
\put(-30,0){\includegraphics[width=0.25\textwidth]{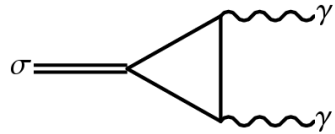}}
\end{picture}}
\caption{ The anomaly diagram exhibiting the coupling of two photons to the (quasiparticle) dilaton. \label{fig:anomaly}}
\end{figure}

With the dynamics of the dilatational current $S^{\mu}$ expressed in the effective $\sigma$ field, inserting the $T_{\mu}^{\mu}$ operator in the quark contribution to the photon self-energy exhibits the scalar-vector-vector anomaly displayed in Fig.~\ref{fig:anomaly}.  Thus low energy photons are produced through the anomaly giving rise to the effective interaction
\beqn\label{Leffconfanom}
\mathcal{L}_{\mathrm{eff}}=\gsgg\sigma F^{\mu\nu}F_{\mu\nu}
\eeqn
where $\gsgg$ is a dimensionful coupling, $F^{\mu\nu}$ is the electromagnetic field tensor. By matching the two-photon decay rate of the $f_0(600)$ meson to the prediction calculated from \req{Leffconfanom}, Basar et al.~\cite{Basar:2012bp} conclude that
\beqn
\gsgg\simeq (50\,\mathrm{GeV})^{-1}
\quad
m_\sigma \simeq 550\,\mathrm{MeV}
\quad
f_\sigma\simeq 100\,\mathrm{MeV}
\eeqn
It must be recognized that these parameters are speculative, and more work is required to understand more precisely the effective coupling of the photons to the $\sigma$, especially viewed as an in-medium (quasi)particle.  Based on these parameters, the low-energy ${\cal L}_{\rm eff}$ \req{Leffconfanom} is valid for the photon momenta $k^{\mu}$ we expect considering the temperature scale attained in the collision $k\sim T\sim 50-90~{\rm MeV}\ll 50~{\rm GeV}$.

The effective interaction [\req{Leffconfanom}] contributes to the photon self-energy, and it is straightforward to calculate the rate of photon production from the imaginary part,  
\begin{align}\label{ImPi}
\mathrm{Im}\,\Pi=\int\!\!& \frac{d^3k_1}{(2\pi)^32\omega_1}\frac{d^3k_2}{(2\pi)^32\omega_2}
\\ \notag &\hspace*{5mm} \times \gsgg^2\!\!\!\sum_{\mathrm{polarizations}}\!\!\!\!(F_1^{\mu\nu}F_2^{\mu\nu})^2\:\mathrm{Im}\,D_\sigma(k_1\!-\!k_2)
%\\ \nolabel
%F_i^{\mu\nu}=k_i^{\mu}\epsilon_i^{\nu}-k_i^{\nu}\epsilon_i^{\mu}
\end{align}
Due to the short mean-free path of photons in medium, we focus on photon emission from the boundary layer (elaborated below) and assume the photon occupation number remains small.  This is a conservative assumption since the statistical factor $1+(\exp(\beta u\cdot k_i)-1)^{-1}$, with $u_{\mu}=(1,\vec 0)$ being the heat-bath 4-velocity, would lead to Bose enhancement of the production rate.
In vacuum, the photon energy $\omega_{1,2}=|\vec k_{1,2}|$, but in medium the dispersion relation is modified. 

The factor $\mathrm{Im}\,D_\sigma(k_1-k_2)$ is the branch cut in the $\sigma$ propagator due to the two-photon continuum.  Explicitly, it is
\beqn
\mathrm{Im}\,D_\sigma(k_1-k_2)=\frac{1}{(m_\sigma^2f_\sigma)^2}\frac{\pi\rho_\sigma(k_1-k_2)}{e^{\beta u\cdot(k_1-k_2)}-1}
\eeqn
with $\beta=1/T$ and the normalization prefactor coming from \req{sigmamatrixelem}.  Like~\cite{Basar:2012bp}, we will for this study consider the in-medium properties of the $\sigma$ to be encapsulated in the spectral function $\rho_\sigma(q^\mu)$, which is related to the bulk properties of the plasma by the Kubo formula [see Eq. 6.155 of~\cite{Kapusta:2006pm}]
\beqn
\lim_{\omega\to 0}\frac{\rho_\sigma(\omega)}{\omega}=\frac{9}{\pi}\zeta
\eeqn
where $\zeta$ is the bulk viscosity.  

The energy and momentum scales of the photons produced by this mechanism are set by $T\lesssim 140~{\rm MeV}$ and the photon spectrum is exponentially suppressed above $k_0\sim T$.  Further, we expect that the momentum dependence of the spectral function $\rho_{\sigma}$ is dominated by QCD physics, and hence varies on a momentum scale given by QCD.  Seeing that $T\lesssim m_{\pi}$ (the lowest QCD momentum scale possible to consider), the photon production will only involve the spectral function near zero frequency in this sense.  Depending on the speed of sound in the dense plasma, it may be interesting to extend our study to see the effect of the sound peak in the spectral function, discussed in~\cite{Meyer:2008gt}.

We consider two cases:
\begin{enumerate}
\item $F^{\mu\nu}_{1,2}$ correspond to two real photons, and hence in \req{ImPi} each $F_i^{\mu\nu}=k_i^{\mu}\epsilon_i^{\nu}-k_i^{\nu}\epsilon_i^{\mu}$ for $i=1,2$.  One must take care to reverse the sign of momentum on the external leg, corresponding to having two outgoing photons in the final state.  In this case, the energy of the two photons is derived from the energy in the collective excitations of plasma, seen in the scalar mode $\sigma$.  Performing the polarization sum, the differential rate for the {\bf diphoton channel} is
\begin{align}\label{2photondiffrate}
\omega_2\frac{d\Gamma_{2\gamma}}{d^3k_2}=\left(\!\frac{\gsgg}{m_\sigma^2f_\sigma}\!\right)^{\!2}\!\!\frac{1}{2\pi^2}\!&\int \!\frac{d^3k_1}{(2\pi)^32\omega_1}
\\ \notag & \times
\frac{\rho_\sigma(\omega_1+\omega_2)}{e^{\beta(\omega_1\!+\!\omega_2)}-1}(2(k_1\!\cdot\! k_2)^2+k_1^2k_2^2)
\end{align}
\item Strong magnetic fields are expected near neutron stars and the external line can be attached to an external magnetic field $F_2^{\mu\nu}=-\epsilon^{\mu\nu\kappa}B_{\kappa}$.  It is for this case that we find it more transparent to set up the calculation as the imaginary part of the photon self-energy.  Since the magnetic field is much more slowly varying than the momentum of the produced photons, the momentum integral $d^3k_2$ is practically a $\delta$ function at zero momentum. In this case, the outgoing photon takes energy from both the collective excitations of the plasma and the external magnetic field.  The differential rate for {\bf $B$-field induced} single photon production is 
\beqn\label{Bdiffrate}
\omega\frac{d\Gamma_{B\gamma}}{d^3k}=\left(\frac{\gsgg}{m_\sigma^2f_\sigma}\right)^2\!\!\frac{1}{2\pi^2}\frac{\rho_\sigma(\omega)}{e^{\beta\omega}-1}(\vec B^2\vec k^2-(\vec B\cdot\vec k)^2)
\eeqn
Note that photons are emitted perpendicular to the axis defined by the $B$-field vector.
\end{enumerate}

%\vskip0.2cm
%%%%%%%%%%%%%%%%%%%%%%%%%%%%%%%%%%%%%%%%%%%%%%%%%%%%%%%%%%%%%%%%%%%%%%%%%
%\section{Plasma Properties and Modified Photon Dispersion}
In a hot dense plasma composed of charged particles, photons have a self-energy due to interactions with the medium.  The photon energy is defined by the solution to 
\begin{align}\label{photondisp}
\omega^2=&\:\vec k^2+G(\omega,\vec k) \\ \notag
\lim_{|\vec k|,\omega\gg T,|\mu_f|}\!\!G(\omega,\vec k)&\equiv \mpl^2 =\frac{1}{2} \sum_f (Q_fe)^2\left(\frac{T^2}{3}+\frac{\mu_f^2}{\pi^2}\right)
\end{align}
$G(\omega,\vec k)$ is the photon self-energy, which can be found evaluated to one fermion loop, e.g. using Eq. (5.51) in~\cite{Kapusta:2006pm}. The sum is over charged particle flavors each having charge $Q_fe$ and chemical potential $\mu_f$.  As indicated, in the high frequency limit, the photons display an effective plasma mass $\mpl$, which sets the scale of the medium effects on photon propagation.
At $2n_s$, $\mpl\simeq 15-20$\,MeV, and we will see the typical photon momentum is $T,\mpl\lesssim|\vec k|< \mu_f$.  The long wavelength excitations $|\vec k|\ll\omega < T,|\mu_f|$ are gapped $\omega^2\simeq \frac{2}{3}\mpl^2+\frac{6}{5}\vec k^2+...$ with the plasma frequency closely related to the plasma mass. 

In numerical integrations in the next section, we solve the photon dispersion for $\omega$ at each value of $|\vec k|$, thus determining the rate of photon production consistently taking into account medium effects.  We note that the peak found in the energy-momentum trace~\cite{Bazavov:2009zn,Basak:2009uv} and bulk viscosity~\cite{Li:2009by} occurs {\it above} the transition temperature and hence in the quark-dominated phase. Therefore, we will evaluate the production rate assuming quarks and electrons are the predominating degrees of freedom where the conformal anomaly and photon production are enhanced; mixed phases may be considered in future work.   The flavor sum $f$ in \req{photondisp} is over $f=u,d,e$ for up and down quarks and electrons. At early times in the collision $t<\tau_{\rm weak}$ before weak interactions have relaxed, the relative numbers of light quarks and electrons remain close to their values as determined by weak equilibrium at zero temperature.   This implies the electron chemical potential is the difference of the down and up quark chemical potentials, $\mu_e=\mu_d-\mu_u$, or equivalently the difference of neutron and proton chemical potentials, $\mu_e=\mu_N-\mu_P$.  The second constraint is charge neutrality, requiring that the numbers of protons and electrons are equal.  We solve these constraints for the proton/neutron ratio at any given density.  The result for protons and neutrons translates into the quark chemical potentials using the fact that the proton and neutron chemical potentials are linear combinations of up and down chemical potentials: $\mu_P=2\mu_u+\mu_d$ and $\mu_N=2\mu_d+\mu_u$.

%%%%%%%%%%%%%%%%%%%%%%%%%%%%%%%%%%%%%%%%%%%%%%%%%%%%%%%%%%%%%%%%
\section{Rate of Energy Loss}
\begin{figure}
\includegraphics[width=0.48\textwidth]{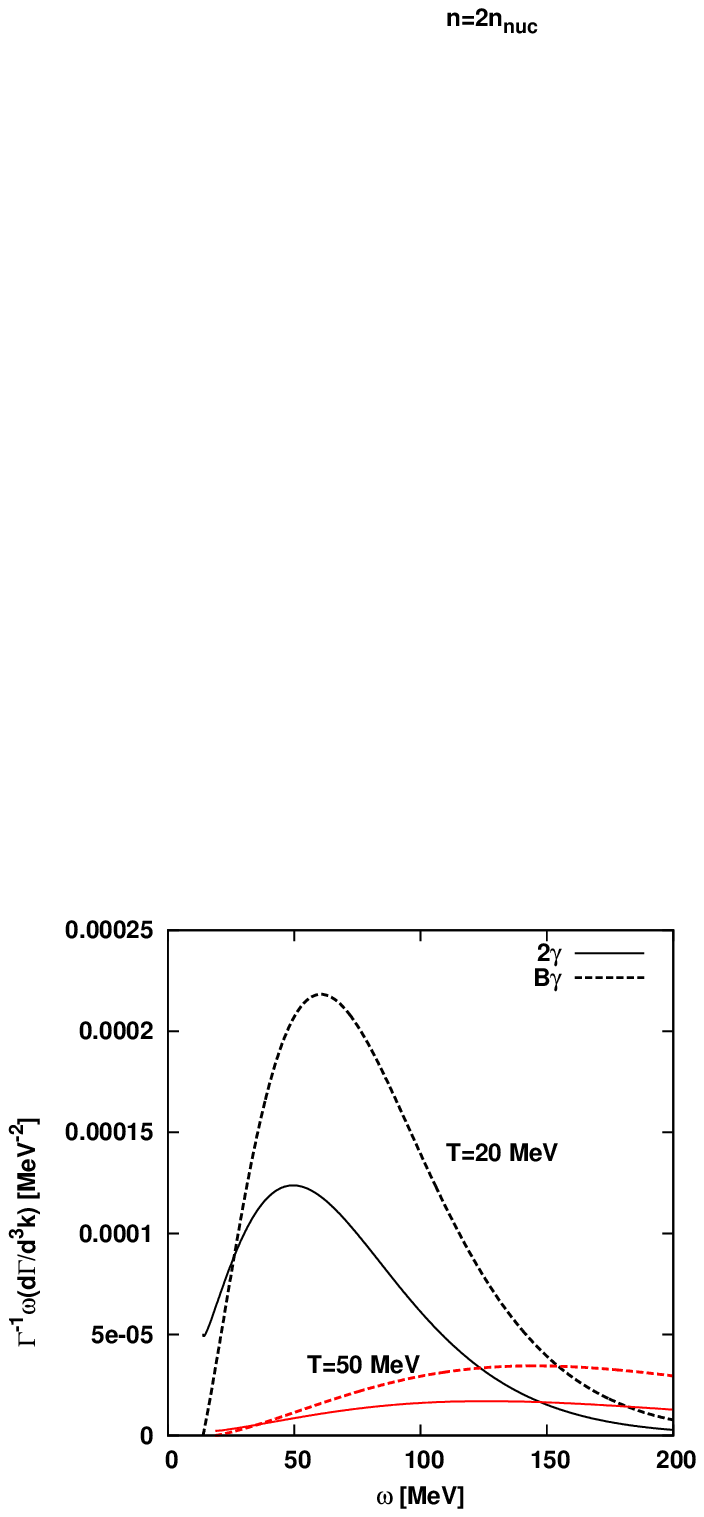}
\caption{ The differential rates [\req{2photondiffrate} and \req{Bdiffrate}], normalized to the total rates $\Gamma_{2\gamma}$,$\Gamma_{B\gamma}$, as a function of photon energy at density $n=2n_{\rm nuc}\simeq 0.34\:{\rm fm}^{-3}$.  Note the gap arising from the photon dispersion relation.  The nonzero value at the lower limit for $2\gamma$ is because this represents the one-photon spectrum, having integrated over the other photon.
\label{fig:spectraTdep}}
\end{figure}

The total number of photons produced is obtained by integrating Eqs.\eqref{2photondiffrate} and \eqref{Bdiffrate} over the photon momentum.  The differential rates Eqs.\eqref{2photondiffrate} and \eqref{Bdiffrate}, normalized to the total rate are plotted in Fig.~\ref{fig:spectraTdep}.  Not all of the volume contributes: due to the high density and finite temperature, the photon has a mean-free path much less than the bulk dimensions of the plasma.  The mean-free path $l_f$ is taken from the imaginary part of the photon self-energy, according to which $l_f\sim 1/e\mpl\simeq 100$\,fm$\,\ll$\,10 m--10km scale of the merging stars.  Photons scatter $N\gg 1$ times before escaping into vacuum.  Considered as a random walk, a photon diffuses a distance $d$ in the time $t=\sqrt{N}d=d^2/l_f$ and would take $3\times 10^6$ years to traverse the 10km radius of the star.  On the other hand, over 100 ns, photons from a depth $\sim 10^7l_f$ are able to diffuse out.

%%%%%%%%%%%%%%%%%%%%%%%%%%%%%%%%%%%%%%%%%%%%%%

We can therefore consider photons as only being emitted from the boundary layer of thickness $\sim l_f$ near where the two phases meet or coexist.  Moreover, only photons with momenta pointing out of the plasma region will escape.  This fact introduces a preferred coordinate system in the evaluation of the rates, oriented relative to the outward pointing normal of the volume element $\hat n$.

To evaluate the momentum carried away by escaping photons, we include a factor of the net momentum $q^{\mu}=k_1^{\mu}+k_2^{\mu}$ before integrating over the photon momenta.  For the $B$-field induced process, $k_2^\mu\to 0$.  In this way, we define the energy and momentum production rates
\begin{align}\label{4momentumrate2gamma}
\Gamma^{\mu}_{2\gamma}&=\!\int\!\!\frac{d^3k_1}{2\omega_1}\frac{d^3k_2}{2\omega_2}(k_1^{\mu}\!+\!k_2^{\mu})\omega_1\omega_2\frac{d\Gamma_{2\gamma}}{d^3k_1d^3k_2} 
\\ \label{4momentumrateBfield}
\Gamma^{\mu}_{B\gamma}&=\!\int\!\frac{d^3k}{2\omega}k^{\mu}\omega\frac{d\Gamma_{B\gamma}}{d^3k} 
\end{align}
Recall that in consideration of the refractive media, the photon frequency is held fixed when transiting from one medium to another.  On exiting the dense matter, the photon 3-momentum in vacuum is determined by $\vec k^2_{\rm vac}=\omega^2$ and hence the zero component $\Gamma^0$ determines the energy emitted from the plasma.

Putting dimensionful quantities into the prefactors, the results are
\begin{align} \label{2gammadE}
\left.\frac{dE}{d^4x}\right|_{2\gamma}\!\!\!&=\left(\frac{\gsgg}{m_\sigma^2f_\sigma}\right)^{\!\!2}\frac{\mpl^{10}}{(2\pi)^3}\frac{9\zeta}{\pi}\:I_{2\gamma}(\beta\mpl)
\\ \notag & 
= 7.13\times 10^3\frac{\mathrm{GeV}}{\mathrm{fm^3 s}}\frac{\zeta}{s}\frac{s}{s_0}\left(\frac{\mpl}{15\,\mathrm{MeV}}\right)^{\!10}I_{2\gamma}(\beta\mpl) \\ 
\label{BfielddE}
\left.\frac{dE}{d^4x}\right|_{B\gamma}\!\!\!&=\left(\frac{\gsgg}{m_\sigma^2f_\sigma}\right)^{\!\!2}\!\frac{2\mpl^6}{3\pi} \frac{9\zeta}{\pi} {\vec B^2}\:I_{B\gamma}(\beta\mpl)
\\ \notag & 
= 5.51\frac{\mathrm{GeV}}{\mathrm{fm^3 s}}\frac{\zeta}{s}\frac{s}{s_0}\left(\frac{\mpl}{15\,\mathrm{MeV}}\right)^{\!6}\frac{\vec B^2}{B_{\mathrm{QED}}^2}I_{B\gamma}(\beta\mpl)
\end{align}
The dimensionless integrals are
\begin{align}\label{I2gamma}
I_{2\gamma}(y)
=&\!\!\int_0^{\infty}\frac{\kappa_1^2d\kappa_1}{w_1}\frac{\kappa_2^2d\kappa_2}{w_2}\frac{(w_1\!+\!w_2)^2}{e^{y(w_1+w_2)}-1}
\\  \notag & \!\times \!\!
\left(\!2w_1^2w_2^2\!-\!\kappa_1\kappa_2w_1w_2\!+\!\frac{2}{3}\kappa_1^2\kappa_2^2\!-\!(w_1^2\!-\!\kappa_1^2)(w_2^2\!-\!\kappa_2^2)\!\right) %(3w_1^2w_2^2+\frac{5}{3}\kappa_1^2\kappa_2^2+\kappa_1\kappa_2w_1w_2-(\kappa_1w_2+\kappa_2w_1)^2)
 \\[2mm] \label{IBfield}
I_{B\gamma}(y)=&\int_0^{\infty}d\kappa\frac{w\kappa^4}{e^{yw}-1}
\end{align}
Here, $\kappa_i$ are dimensionless integration variables, named to recall that they are scaled $\kappa_i =|k_i^\mu|/\mpl$ and $w_i=\omega_i/\mpl$ from the momentum and energy appearing in the photon phase space integrals in Eqs.\,\eqref{4momentumrate2gamma} and \eqref{4momentumrateBfield}.  Recall that $\omega_i$ is determined from the dispersion relation \req{photondisp}.

We have scaled the bulk viscosity by the entropy density $s$, because the  ratio $\zeta/s$ is shown to be approximately constant as a function of density and temperature in the weakly coupled quark-gluon plasma~\cite{Chen:2012jc}.  $\zeta/s$ has recently been the subject of much study as a characterisation of the strongly interacting plasma, see for example~\cite{Kharzeev:2007wb,Karsch:2007jc,Chen:2011km}, and in some studies performed at zero density~\cite{Kharzeev:2007wb,Karsch:2007jc} found to be increasing towards order $\zeta/s\sim 1$ as the phase transition temperature is approached from above. To see the quantitative scale, the entropy density is normalized by the entropy density of a noninteracting up+down quark-electron plasma at $T=50$~MeV and $n=2n_{\rm nuc}$. Note that we can reasonably expect $\zeta$ to be controlled by QCD interactions because the relaxation time scales associated with this photon emission are generally faster than the weak interaction relaxation times.  The frequency of the perturbation to the plasma is therefore faster than weak interactions can respond and we can consistently neglect weak interaction dynamics in this study.  However, especially considering domains near to phase transition, little is known about the numerical value of $\zeta/s$, which can be very different from order unity.  $B_{\mathrm{QED}}=m_e^2/e=4.41\times 10^9$\,T is the QED field scale.  Note that the diphoton channel dominates over the $B$-field assisted channel unless $B\gtrsim 100B_\mathrm{QED}$.

%%%%%%%%%%%%%%%%%%%%%%%%%%%%%%%%%%%%%%%%%%%%%%
\begin{figure}
\includegraphics[width=0.48\textwidth]{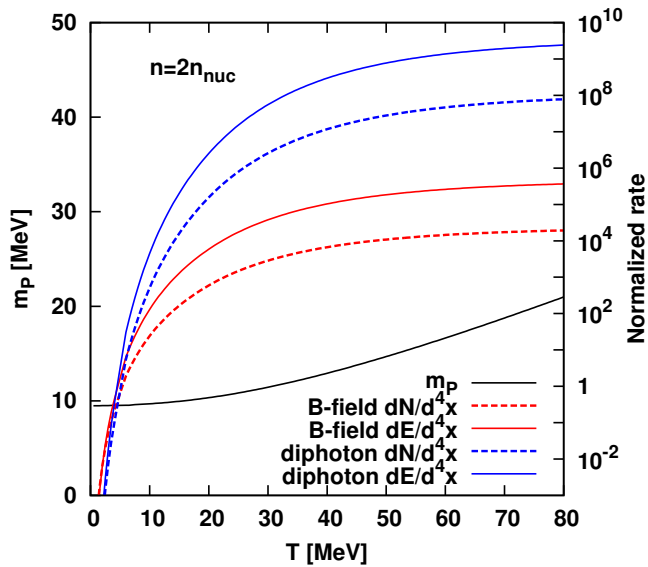}
\caption{ At fixed density $2n_{\rm nuc}$, the plasma mass in \req{photondisp} (solid black line) with units on left axis.  The rates of photon number and energy emission are plotted on the right axis, normalized by removing the dimensionful prefactor.  For energy, this corresponds to plotting the integrals $I_{2\gamma}$ \req{I2gamma} (solid) and $I_{B\gamma}$ \req{IBfield} (dashed), which at fixed density depend only on temperature.
\label{fig:rateIfns}  }
\end{figure}
%%%%%%%%%%%%%%%%%%%%%%%%%%%%%%%%%%%%%%%%%%%%%% 

The photons are emitted with an energy of the order of the temperature and plasma properties, i.e. tens of MeV.  This energy scale is significantly higher than the photon energy typically discussed in the context of gamma ray bursts.  On the other hand, these photons constitute an ``ultraprompt'' signal that may escape without interaction in the first tens of nanoseconds.  Bremsstrahlung radiation and Compton scattering also contribute to the spectrum of photons emitted.  From the perspective of directly detecting the photons produced by the processes considered here, these sources are a background that would have to be studied in greater detail.  The characteristic energy scales associated with this process may provide a feature allowing us to separate the signal from these backgrounds.

Over the duration of the merger, the environment near the stars is cluttered by plasma and ejecta.  Compton scattering in this plasma would be likely to soften the spectrum of later emission, but modeling this effect is beyond the scope of the present work.  Another effect we have not been able to consider in the present framework is multiple $2,4,6...$-photon production.  It is recognized in nonperturbative studies of the conformal anomaly~\cite{Labun:2008qq}, that the energy-momentum trace induces effective couplings $2N$, $N\geq 1$ photons.  In the presence of the strong fluctuations near the phase transition and strongly coupled quark matter in this density-temperature domain, the $N>1$ effective couplings can be important.

%\vskip0.2cm
%%%%%%%%%%%%%%%%%%%%%%%%%%%%%%%%%%%%%%%%%%%%%%%%%%%%%%%%%%%%%%%%%%%%%%%%%
\section{Cooling time scale} 
The energy loss from the surface layer leads to cooling of the domain of quark matter to a depth of $\sim l_f=100~{\rm fm}$.  If enough energy is released that the temperature drops below the transition region, photon emission is reduced, because the conformal anomaly and bulk viscosity are no longer near their peaks.  On the other hand, if heat from the bulk is conducted to the surface layer sufficiently quickly, then the cooling of the surface is compensated and energy escapes continuously from the whole quark matter domain at the rate determined by \req{2gammadE}.  To characterise this rate, we introduce the time scale
\begin{align}\label{tauE}
\frac{1}{\tau_E}&=\frac{1}{\varepsilon}\left.\frac{dE}{dVdt}\right|_{2\gamma}\!\!=
\left(\frac{\gsgg}{m_\sigma^2f_\sigma}\right)^{\!\!2}\frac{9\mpl^{10}}{8\pi^4}\frac{\zeta}{s}\frac{s}{\varepsilon}I_{2\gamma}(\beta\mpl)  
\\ \notag  
&= (1.0\times 10^{-6}\,{\rm s})^{-1}\frac{\zeta}{s}\left(\frac{\mpl}{15\,\mathrm{MeV}}\right)^{10}\frac{s/\varepsilon}{(s/\varepsilon)_0}I_{2\gamma}(\beta\mpl)  
\end{align}
where the entropy to energy ratio $(s/\varepsilon)_0$ has been evaluated at $T=50~{\rm MeV}$ and $n=2n_{\rm nuc}$.  Note from Fig.~\ref{fig:rateIfns} that at even moderate temperatures, the integrals $I_{2\gamma},I_{B\gamma}$ are large multiplicative factors.  At $T=50\,{\rm MeV}$, $\tau_E$ is on the order of $(\zeta/s)^{-1}10^{-15}~{\rm s}$, much less than the equilibration time scale for weak interactions.

The rate at which heat is transported to the surface layer is controlled by the thermal conductivity of the quark matter.   For a baseline comparison, we take the perturbative estimate of the thermal conductivity, such as that found in~\cite{Heiselberg:1993cr} and define the time scale for diffusion over a length $R$ as
\beqn\label{tauk}
\tau_\kappa = \frac{c_VR^2}{\kappa}=2\times 10^{-21}\,{\rm s}\left(\frac{R}{100~{\rm fm}}\right)^2
\eeqn
in which $c_V\simeq \pi^2\sum_fn_fT/\mu_f$ is the specific heat and the thermal conductivity goes as $\kappa\simeq \sum_f\mu_f^2/\alpha_s$.  The numerical coefficient is obtained for the case $T\ll 2\mu_f$, see Eq. (62) of~\cite{Heiselberg:1993cr}.

These time scales show sensitivity to properties of the quark matter, the thermal conductivity $\kappa$ and bulk viscosity $\zeta$.  Their numerical prefactors may not reflect the true situation: $\zeta/s$ can take on values ranging across many orders of magnitude, and little is known about how it behaves in the specific domain of interest near a phase transition at high density.  The thermal conductivity also needs careful attention.

The situation becomes especially interesting if the time scales [\req{tauE} and \req{tauk}] turn out to be commensurate: in this case, there can be a slow $\tau\gg\tau_E,\tau_\kappa$ cooling.  If the cooling is sufficient to drop out of the transition region, photon emission is suppressed, because the enhancement of the conformal anomaly is no longer present and the mean-free path of photons in the hadron phase can be up to an order of magnitude shorter than in the quark phase.  Such a ``darkening'' phenomenon would produce characteristic temporal signatures in the spectrum of electromagnetic radiation obtained from bNS mergers.

%%%%%%%%%%%%%%%%%%%%%%%%%%%%%%%%%%%%%%%%%%%%%%%%%%%%%%%%%%%%%%%%%%%%%%%%%
\section{Conclusions}

In simplified analytic models of a binary neutron star merger (noting also the work of Shapiro~\cite{Shapiro:1998sy}), we showed that matter in the collision achieves densities and temperatures in a range where a transition to quark degrees of freedom can be expected.  We have identified photon production from the conformal anomaly as a signal of the transition and probe of the properties of the quark matter and the environment during the collision.  As seen in Eqs.\,\eqref{2gammadE} and \eqref{BfielddE}, the overall magnitude of photon production and energy loss depend on the bulk viscosity of the dense matter and magnetic fields present during the collision.  The thermal conductivity enters in consideration of the net cooling effect of the energy loss.

The properties of matter in this region of the phase diagram remain a challenging subject.  Lattice calculations have recently made progress by studying fixed baryon number in the canonical ensemble~\cite{Li:2011ee}, and one may look forward to investigations of the relevant thermodynamic and transport properties as input in studies like this one aiming to understand signals from bNS mergers.  There may also be an opportunity to distinguish different structures in this region of the QCD phase diagram (also including magnetic field effects), using for example distinct signatures of the quarkyonic phase~\cite{Torrieri:2013mk}.

Combining these other studies, we may find phenomenology characteristic to the transition to quark matter during a bNS merger and thus potentially forge a (nongravitational) observational connection between bNS and short gamma ray bursts.  For example, the cooling effect could lead to the interesting possibility of ``surface darkening'', which would be imprinted in the time variation of the electromagnetic radiation obtained from the bNS.  Such an effect was anticipated in~\cite{Chen:2013tp}, and a specific study of the energy balance and rates is an important next step in this direction. The application of our results to the phenomenology of sGRBs will be the subject of future work.

Our result for the cooling time scale, \req{tauE}, indicates that dynamics of a possible QCD transition remain faster than weak interactions.  This shows that our study is consistent in considering the flavor content as fixed and weak reactions as a higher order correction.  Seeing that weak reactions do become important over the course of the merger [recall \req{times}] and the flavor content will evolve, we will return to the study of the influence of strange quarks in future work.  The properties of the 3-flavor quark matter, with weak interactions active, will be very different and is a clear next step in investigation.

Extending this study both toward phenomenology at later times in the collision and to include other signals, particularly neutrinos, is essential to improve our ability to compare to observations.
The physics involved presents both a challenge and an opportunity: the properties of the dense matter are not yet well known, and our evaluations are intended to provide a baseline for future work and the beginning of comparison to observations.
 
%\vskip 0.5cm
%%%%%%%%%%%%%%%%%%%%%%%%%%%%%%%%%%%%%%%%%%%%%%%%%%%%%%%%%%%%%%%%
{\it Acknowledgments:}  We thank J.-W. Chen, M. Chernodub, M. Huang, A. Schmitt, and I. Shovkovy for discussion.  We appreciate the input of K.-F. Liu in discussions and reading an early draft of this manuscript.
Pisin Chen is supported by Taiwan National Science Council under Project No. NSC 97-2112-M-002-026-MY3 and by US Department of Energy under Contract No. DE-AC03-76SF00515.
%%%%%%%%%%%%%%%%%%%%%%%%%%%%%%%%%%%%%%%%%%%%%%%%%%%%%%%%%%%%%%%%

\end{document}